\documentstyle[12pt,a4,fleqn]{article}

\title{Quantum interference of polarized electrons\\
in the presence of magnetic strings}
\author{Silviu Olariu\\
Institute of Physics and Nuclear Engineering, Heavy-Ion Physics Department\\
76900 Magurele, P.O. Box MG-6, Bucharest, Romania\\}

\begin{document}

\date{}
\maketitle

\abstract

The interaction of polarized electrons with bare and shielded magnetic strings 
is studied  with the aid of the Dirac equation.
It is found that the difference between the amplitudes for the scattering by 
bare and shielded strings of incident wave packets of width $\delta$ and 
impact parameter $d$  is proportional to $\exp(-d^2/2\delta^2)$.\\

\endabstract


\setlength{\parindent}{0cm}

\newpage

The basic requirement of an Aharonov-Bohm experiment \cite{1} is the separation
between the region of space accessible to the interfering electron and the
region of space containing the magnetic flux. In electron interference 
experiments this separation is obtained by placing suitable shields in the
path of the incident electron beam. In experiments with Josephson junctions
or mesoscopic rings the electrons are prevented from leaving the conductors
by surface potential barriers. In theoretical models the Aharonov-Bohm effect
is demonstrated by showing the persistence of finite, flux-dependent phase
shifts in the presence of barriers of arbitrary height or thickness.\\

Percoco and Villalba \cite{2}  have studied the eigenfunctions of the Dirac
equation in the presence of a magnetic string shielded by a cylindrical
barrier and have found that for a barrier of finite radius it is
not possible that all four components of the Dirac wave function should vanish
on the surface of this cylinder. Therefore, when we represent the state of the
electron by solutions of the Dirac equation, the separation between the region
accessible to the electron and the region of the magnetic flux has to be
achieved by a barrier of finite height and sufficient thickness.\\

A flux of $h/2e=2.07\cdot10^{-15}$ Tm$^2$ which would produce a shift of
a half-fringe in an electron interference pattern would be obtained from
a 2 T magnetic field from a tube of magnetic flux having a radius 
$r_{h/2e}=1.81\cdot10^{-8}$ m, and the wave vector of an electron having an 
energy of 1 keV is $\tilde{k}_1=1.62\cdot10^{11}$ m$^{-1}$, so that
$\tilde{k}_1r_{h/2e}=2.9\cdot10^3\gg 1$. Nevertheless, a concept often used in 
theoretical models is that of magnetic string, which is a tube of arbitrarily
small radius carrying a certain {\it finite} amount of magnetic flux $F$. The
interaction of a particle of charge $q$ with the flux $F$ is measured by the
coupling constant $\alpha=qF/2\pi\hbar$, in Syst\`{e}me International (SI)
 units. For an electron we have $q=-e<0$.\\

If the state of the particle of charge $q$ is represented by the 
Schr\"{o}dinger equation, thereby neglecting the electron spin, the 
eigenfunctions in the presence of a magnetic string are all vanishing on the
string line, so that the inclusion of additional shielding barriers is not
necessary in this case. A finite amount of momentum is exchanged, however,
between the electron and the string at the surface of the string \cite{3}.
The eigenfunctions of an electron endowed with spin interacting with a magnetic
string may in some cases differ significantly from the eigenfuctions of a 
spinless electron interacting with that magnetic string. Hagen
\cite{4} and Olariu \cite{5} have found that the representation of the electron
states by solutions of the Dirac equation involves an eigenfunction which
becomes infinite on the magnetic string. However, while \cite{4} obtained
divergent terms in the pair of principal components of the four-component
Dirac wave function, in \cite{5} the divergent terms were proportional to
$\hbar k/Mc$, becoming vanishingly small in the  limit $\hbar k/Mc
\rightarrow 0$, where $\hbar k$ is the momentum and $M$ the mass of the 
incident electron.\\

The Dirac eigenfunctions of an electron in the presence of a shielded
magnetic string are determined in this work  by a limiting process in which
the radius $r_0$ of the tube of magnetic flux tends to zero and the product 
$kR_0$ tends to zero while the radius $R_0$ of the shielding barrier remains 
finite.
It is shown that the scattering of polarized electrons by a shielded magnetic
string can be represented, in the limit $\hbar k/Mc \rightarrow 0$, by the
Aharonov-Bohm wave function for spinless electrons.
The scattering of a plane wave by a shielded magnetic string is then compared 
with the scattering of a plane wave by a bare magnetic string, and it is shown
that the difference between the scattering by bare and shielded strings
persists in the non-relativistic limit.
By comparing
further the scattering of wave {\it packets} by bare and shielded magnetic 
strings it is found that the difference of the corresponding scattering 
amplitudes is 
proportional to the probability density in the region of the string. This
probability is made as small as possible in real Aharonov-Bohm experiments.
Thus, the scattering by a shielded magnetic string and the scattering
by a bare magnetic string are really different only when there is a significant
probability density in the region of the magnetic flux, i.e. when there is
a direct interaction between the magnetic moment of the electron and the
magnetic {\it field}.   \\

The separation between the region of space accessible to the incident electron
and the region of the magnetic flux is a basic requirement of an Aharonov-Bohm
experiment. The concept of {\it shielded} magnetic string has been introduced
\cite{3},\cite{5} by assuming that the tube of magnetic flux of radius $r_0$
is surrounded by a cylindrical potential barrrier of height $U$, 
$E-Mc^2<U<E+Mc^2$, and radius $R_0>r_0$, where $E$ is the electron
energy, including rest mass. The eigenfunctions for a shielded magnetig 
{\it string} can then be determined for fixed $R_0$ in the limit 
$r_0\rightarrow0$ and $kR_0\rightarrow0$, the latter being equivalent to
$k\rightarrow 0$. Thus, in addition to the vector potential of the tube of 
magnetic flux, of radius $r_0$, which in cylindrical coordinates is
\begin{eqnarray}
A_z^{(r_0)}&=&0, \nonumber\\ \:A_r^{(r_0)}&=&0,\nonumber\\
A_\theta^{(r_0)}&=&\left\{\begin{array}{rcl} F/2\pi r & {\rm for} &
r>r_0,\\ Fr/2\pi r_0^2 & {\rm for} & r<r_0 , \end{array}\right.
\end{eqnarray}
we assume that there is a static electric potential given by
\begin{equation}
q\phi=\left\{ \begin{array}{ccc} U & {\rm for} & r<R_0,\\ 0 & 
{\rm for} & r>R_0 ,
\end{array}\right.
\end{equation}
where $q$ is the charge of the incident particle and $r$ the distance to
the axis of the tube of flux.\\

The eigenstates of the particle of charge $q$ in the presence
of the potentials represented in eqs. (1) and (2) have the form
\begin{equation}
\left(\begin{array}{c}\Psi_{l1}^{(R_0)}\\ \Psi_{l2}^{(R_0)}\end{array}\right)=
\left(\begin{array}{c}\chi_{l1}^{(R_0)}(r)e^{il\theta}\\ 
\chi_{l2}^{(R_0)}(r)e^{i(l+1)\theta} 
\end{array}\right)e^{-iEt/\hbar},
\left(\begin{array}{c}\Psi_{l3}^{(R_0)}\\ \Psi_{l4}^{(R_0)}\end{array}\right)=
\left(\begin{array}{c}\chi_{l3}^{(R_0)}(r)e^{il\theta}\\ 
\chi_{l4}^{(R_0)}(r)e^{i(l+1)\theta} 
\end{array}\right)e^{-iEt/\hbar},
\end{equation}
where $l$ is an integer. The radial parts are determined in the standard
representation \cite{1}, \cite{5a} and in SI units, from the relations
\begin{equation}
\frac{d^2\chi_{l1}^{(R_0)}}{d r^2}+\frac{1}{r}\frac{d
\chi_{l1}^{(R_0)}}{d r}+
\left[\frac{(E-q\phi(r))^2-M^2c^4}{\hbar^2c^2}-\frac{1}{r^2}
\left(l-\frac{q}{\hbar}rA_\theta^{(r_0)}(r)\right)^2+\frac{q}{\hbar}
B(r)\right]\chi_{l1}^{(R_0)}=0,
\end{equation}
\begin{equation}
\frac{d^2\chi_{l2}^{(R_0)}}{d r^2}+\frac{1}{r}\frac{d
\chi_{l2}^{(R_0)}}{d r}+
\left[\frac{(E-q\phi(r))^2-M^2c^4}{\hbar^2c^2}-\frac{1}{r^2}
\left(l+1-\frac{q}{\hbar}rA_\theta^{(r_0)}(r)\right)^2-\frac{q}{\hbar}
B(r)\right]\chi_{l2}^{(R_0)}=0,
\end{equation}
\begin{equation}
\chi_{l3}^{(R_0)}=-\frac{i\hbar c}{E+Mc^2-q\phi(r)}\left(\frac{d}{d r}
+\frac{l+1}{r}-\frac{q}{\hbar}A_\theta^{(r_0)}(r)\right)\chi_{l2}^{(R_0)},
\end{equation}
\begin{equation}
\chi_{l4}^{(R_0)}=-\frac{i\hbar c}{E+Mc^2-q\phi(r)}\left(\frac{d}{d r}
-\frac{l}{r}+\frac{q}{\hbar}A_\theta^{(r_0)}(r)\right)\chi_{l1}^{(R_0)},
\end{equation}
where $B(r)$ is the magnetic field represented by the vector potential (1).\\

In the region of 
the magnetic flux, $r<r_0$, the radial eigenfunctions are determined by the 
condition
of regularity at $r$=0, and are 
\begin{equation}
\chi_{l1}^{(R_0)}=a_{l1}B_{l1}^{(R_0)}(\kappa_1 r)^{|l|}
\exp(-\alpha r^2/2r_0^2)
F\left(\left.\frac{|l|+1-l}{2}-\frac{\kappa_1^2 r_0^2}{4\alpha}
\right | |l|+1\left |
\frac{\alpha r^2}{r_0^2}\right.\right) \:\,{\rm for}\:\ r<r_0,
\end{equation} 
\begin{equation}
\chi_{l2}^{(R_0)}=a_{l2}B_{l2}^{(R_0)}(\kappa_2 r)^{|l+1|}
\exp(-\alpha r^2/2r_0^2)
F\left(\left.\frac{|l+1|-l}{2}-\frac{\kappa_2^2 r_0^2}{4\alpha}
\right | |l+1|+1\left |
\frac{\alpha r^2}{r_0^2}\right.\right) \:\,{\rm for}\:\, r<r_0,
\end{equation}
where $a_{l1}, a_{l2}, B_{l1}^{(R_0)}, B_{l2}^{(R_0)}$ are constant 
coefficients, $ B=F/\pi r_0^2$, 
\begin{equation}
\kappa_1^2=-\kappa^2+\frac{qB}{\hbar}, \:\:  \kappa_2^2=
-\kappa^2-\frac{qB}{\hbar},
\end{equation}
\begin{equation}
-\kappa^2=\frac{(E-U)^2-M^2c^4}{\hbar^2c^2}<0 ,
\end{equation}
and $F(a|c|z)$ is the confluent hypergeometric
function \cite{5b} .\\

The eigenfunctions for a bare magnetic string will be obtained first as the
eigenfunctions for a tube of magnetic flux of radius $r_0$,  in the limit
$r_0\rightarrow0$. The radial solutions $\chi_{l1}, \chi_{l2}$ 
of the wave equations for the flux region
$r<r_0$ can be obtained from eqs. (8), (9)
by putting  $U=0$ in eq. (11). In the field-free region $r>r_0$, the radial 
solutions
are of the form 
\begin{equation}
\chi_{l1}=a_{l1}\left(J_{|l-\alpha|}(kr)+A_{l1}H_{|l-\alpha|}
^{(1)}(kr)\right) \:\,{\rm for}\:\, r>r_0,
\end{equation}
\begin{equation}
\chi_{l2}=a_{l2}\left(J_{|l+1-\alpha|}(kr)+A_{l2}H_{|l+1-\alpha|}
^{(1)}(kr)\right) \:\,{\rm for}\:\, r>r_0 ,
\end{equation}
where
\begin{equation}
k^2=\frac{E^2-M^2c^4}{\hbar^2c^2} .
\end{equation}
The coefficients $A_{l1},A_{l2}$ can be determined from the conditions of
continuity at $r=r_0$ of the four components of the wave function, and have
the expressions 
\begin{equation}
A_{l1}=-\frac{J_{|l-\alpha|}^\prime(kr_0)-\Lambda_1(k_1/k)J_{|l-\alpha|}(kr_0)}
{H_{|l-\alpha|}^{(1)\prime}(kr_0)-\Lambda_1(k_1/k)H_{|l-\alpha|}^{(1)}(kr_0)} ,
\end{equation}
\begin{equation}
A_{l2}=-\frac{J_{|l+1-\alpha|}^\prime(kr_0)-\Lambda_2(k_2/k)J_{|l+1-\alpha|}
(kr_0)}{H_{|l+1-\alpha|}^{(1)\prime}(kr_0)-\Lambda_2(k_2/k)H_{|l+1-\alpha|}^
{(1)}(kr_0)} ,
\end{equation}
where  $J_\nu^\prime(\zeta)=dJ_\nu(\zeta)/d\zeta, \: H_\nu^{(1)\prime}(\zeta)
=dH_\nu^{(1)}(\zeta)/d\zeta ; \: \Lambda_1=(d\chi_{l1}/k_1dr)/\chi_{l1}$,
$\Lambda_2=(d\chi_{l2}/k_2dr)/\chi_{l2}$ are the logarithmic derivatives 
at $r=r_0$ for the  region $r<r_0$; and $k_1^2=k^2+qB/\hbar,
k_2^2=k^2-qB/\hbar$.\\

As can be seen from eqs. (8), (9) the logarithmic derivative $\Lambda_1$
is, for $l\ge 0$ and in the limit $r_0\rightarrow 0$, 
$\Lambda_1\rightarrow (|l|-\alpha)/
k_1r_0$. 
For $kr_0 \ll 1$, the leading term of the denominator in eq. (15) is 
proportional to
\begin{equation}
H_{|l-\alpha|}^{(1)\prime}(kr_0)-\Lambda_1(k_1/k)H_{|l-\alpha|}^{(1)}(kr_0) 
\sim |l-\alpha|+|l|-\alpha .
\end{equation}
When $kr_0\ll 1$, the
coefficients are $A_{l1} \sim (kr_0)^{2|l-\alpha|}, $
becoming vanishingly small unless
\begin{equation}
|l-\alpha|+|l|-\alpha=0 , \:\: |l-\alpha|<1 ,
\end{equation}
when the ratio in eq. (15) tends to a finite limit.
This equation has solutions only when $\alpha>0$, the solution being
$l=[\alpha]$, where $[\alpha]$ is the integer part of $\alpha$, i.e. the 
nearest integer
to the left of $\alpha$, so that $[\alpha]\le \alpha<[\alpha]+1$. 
It can be shown with the aid of the relation 
$H_\nu^{(1)}=(i/\sin(\pi\nu))\left(e^{-i\pi\nu}J_\nu-J_{-\nu}\right)$ \cite{5c}
that $A_{[\alpha]1} \rightarrow i\sin(\pi\alpha)e^{i\pi\alpha}$ , 
so that $\chi_{[\alpha]1}\sim J_{-\alpha+[\alpha]}(kr)$, instead of the usual 
$J_{\alpha-[\alpha]}(kr)$. For $l<0, \Lambda_1$ remains of the order of
$1/k_1 r_0$, but the leading term of the denominator in eq. (15) is different
from 0 for all values of $l$ in this range, and the coefficients 
$A_{l1} \sim (kr_0)^{2|l-\alpha|}, $
are becoming vanishingly small as $kr_0\ll 1$.
The logarithmic derivative $\Lambda_2$ is, for $l\le -1$
and in the limit $r_0\rightarrow 0$, 
$\Lambda_2\rightarrow (|l+1|+\alpha)/k_2r_0$. 
For $kr_0 \ll 1$, the leading term of the denominator in eq. (16) is 
proportional to
\begin{equation}
H_{|l+1-\alpha|}^{(1)\prime}(kr_0)-\Lambda_2(k_2/k)H_{|l+1-\alpha|}^
{(1)}(kr_0) \sim |l+1-\alpha|+|l+1|+\alpha ,
\end{equation}
and when $kr_0\ll 1$, the
coefficients are $ A_{l2} \sim (kr_0)^
{2|l+1-\alpha|}$, becoming vanishingly small, unless
\begin{equation}
|l+1-\alpha|+|l+1|+\alpha=0 ,\:\:|l+1-\alpha|<1 ,
\end{equation}
when the ratio in eq. (16) tends to a finite limit.
This equation has solutions only when $\alpha<0$, the solution being again
$l=[\alpha]$. It can be shown  that
$A_{[\alpha]2}\rightarrow -i\sin(\pi\alpha)e^{-i\pi\alpha}$ , 
so that $\chi_{[\alpha]2}\sim J_{\alpha-[\alpha]-1}(kr)$, instead of the 
usual $J_{[\alpha]+1-\alpha}(kr)$. For $l> -1, \Lambda_2$ remains of the order
of $1/k_2 r_0$, but the leading term of the denominator in eq. (16)
is different from 0 for all values of $l$ in this range, and the coefficients 
are $ A_{l2} \sim (kr_0)^
{2|l+1-\alpha|}$, becoming vanishingly small as $kr_0\ll 1$.
Thus,  in the limit $r_0\rightarrow 0$ 
the radial eigenfunctions for the bare string are, in the region $r>r_0$ and 
for $\alpha>0$, 
\begin{equation}
\chi_{l1}\sim J_{|l-\alpha|}(k r) , \: l\not=[\alpha];\:
\chi_{[\alpha]1}\sim J_{-\alpha+[\alpha]}(k r) \:{\rm for}\: 
\alpha>0 ,
\end{equation}
\begin{equation}
\chi_{l2}\sim J_{|l+1-\alpha|}(k r) \:{\rm for} \: \alpha>0 ,
\end{equation}
and, for $\alpha<0$,
\begin{equation}
\chi_{l1}\sim J_{|l-\alpha|}(k r) \: {\rm for} \: \alpha<0 ,
\end{equation}
\begin{equation}
\chi_{l2}\sim J_{|l+1-\alpha|}(k r), \: l\not=[\alpha];\:
\chi_{[\alpha]2}\sim J_{\alpha-[\alpha]-1}(k r) 
\:{\rm for} \:\alpha<0 .
\end{equation}

It can be checked that the cancellations occur
when the magnetic moment of the electron and the magnetic flux have the same
orientation. For example, in the case of an electron of charge $q=-e<0$ a
positive $\alpha$ means a negative $F$, i.e. a magnetic flux oriented in the
-z direction. For $\alpha>0$ the cancellation occurs for the 
$\chi_{[\alpha]1}$ component,
which represents a spin oriented in the +z direction, so that the magnetic
moment is oriented in the -z direction, as the flux $F$.\\

The solutions for a shielding barrier of fixed radius $R_0$ enclosing a 
flux tube of radius $r_0\rightarrow 0$ will be discussed further.
In the shielding region $r_0<r<R_0$ the  radial eigenfunctions
are superpositions of Bessel functions of imaginary argument $i\kappa r$, 
\begin{equation}
\chi_{l1}^{(R_0)}=a_{l1}\left(C_{l1}^{(R_0)}H_{|l-\alpha|}^{(1)}(i\kappa r)
+D_{l1}^{(R_0)}H_{|l-\alpha|}^{(2)}(i\kappa r)\right)
\:\,{\rm for}\:\, r_0<r<R_0,
\end{equation}
\begin{equation}
\chi_{l2}^{(R_0)}=a_{l2}\left(C_{l2}^{(R_0)}H_{|l+1-\alpha|}^{(1)}(i\kappa r)
+D_{l2}^{(R_0)}H_{|l+1-\alpha|}^{(2)}(i\kappa r)\right)
\:\,{\rm for}\:\, r_0<r<R_0,
\end{equation}
where $C_{l1}^{(R_0)}, D_{l1}^{(R_0)}, C_{l2}^{(R_0)}, D_{l2}^{(R_0)}$ are 
constant coefficients, and
$H_\nu^{(1)}, H_\nu^{(2)}$ are the Hankel functions of the first and second 
kind \cite{5c}.
It can be shown as previously that in the limit $r_0\rightarrow 0$ 
the radial eigenfunctions in the region $r_0<r<R_0$ are, for $\alpha>0$, 
\begin{equation}
\chi_{l1}^{(R_0)}\sim J_{|l-\alpha|}(i\kappa r) , \: l\not=[\alpha];\:
\chi_{[\alpha]1}^{(R_0)}\sim J_{-\alpha+[\alpha]}(i\kappa r) \:{\rm for}\: 
\alpha>0 ,
\end{equation}
\begin{equation}
\chi_{l2}^{(R_0)}\sim J_{|l+1-\alpha|}(i\kappa r) \:{\rm for} \: \alpha>0 ,
\end{equation}
and, for $\alpha<0$,
\begin{equation}
\chi_{l1}^{(R_0)}\sim J_{|l-\alpha|}(i\kappa r) \: {\rm for} \: \alpha<0 ,
\end{equation}
\begin{equation}
\chi_{l2}^{(R_0)}\sim J_{|l+1-\alpha|}(i\kappa r), \: l\not=[\alpha];\:
\chi_{[\alpha]2}^{(R_0)}\sim J_{\alpha-[\alpha]-1}(i\kappa r) 
\:{\rm for} \:\alpha<0 .
\end{equation}
As can be seen from eqs. (21)-(24) and (27)-(30), the presence of the 
barrier changes the argument of the Bessel solutions from a real number in the
case of the bare string into an imaginary number in the case of the shielded 
string, but does not change the divergent character of one of the principal
components of the eigenfunction for $l=[\alpha]$ at $r=r_0$,
when $r_0\rightarrow 0$. However, if $\kappa R_0\gg 1$,   
{\it all} the eigenfunctions in eqs. (27), (28) or (29), (30) become for
$r<R_0$ and $r\approx R_0$  proportional to $\exp(\kappa r)/r^{1/2}$. 
Therefore, the eigenfunctions  are multiplied in the region $r<R_0$
by  coefficients of the order of $\exp(-\kappa R_0)$, which
become vanishingly small for $\kappa R_0 \gg 1$.
For example, if we would choose, in a gedanken experiment, $U=Mc^2$, then
$\kappa\approx Mc/\hbar$, and we would have to fulfil the conditions 
$McR_0/\hbar\gg1$ (effective shielding) and $kR_0\ll 1$ (thin flux tube).
For $k=1.62\cdot 10^{11}$ m$^{-1}$, which corresponds to a kinetic energy
$\hbar^2k^2/2M$=1 keV, and $\hbar/Mc=3.86\cdot 10^{-13}$ m, we could take
$R_0=10^{-12}$ m, so that $kR_0$=0.16, $McR_0/\hbar$=2.59, and 
$\exp(-2McR_0/\hbar)=5.6\cdot 10^{-3}$.\\

In the region $r>R_0$ the radial principal components $\chi_{l1}^{(R_0)},
\chi_{l2}^{(R_0)}$ have the form
\begin{equation}
\chi_{l1}^{(R_0)}=a_{l1}\left(J_{|l-\alpha|}(kr)+A_{l1}^{(R_0)}H_{|l-\alpha|}
^{(1)}(kr)\right) \:\,{\rm for}\:\,r>R_0,
\end{equation}
\begin{equation}
\chi_{l2}^{(R_0)}=a_{l2}\left(J_{|l+1-\alpha|}(kr)+A_{l2}^{(R_0)}H_{|l+1
-\alpha|}^{(1)}(kr)\right) \:\,{\rm for}\:\,r>R_0,
\end{equation}
The coefficients $A_{l1}^{(R_0)}, A_{l2}^{(R_0)}$ appearing in the expression
of the eigenfunctions for $r>R_0$, eqs. (31),(32), can be determined from the
condition of continuity at $r=R_0$ of the four components $\chi_{l1}^{(R_0)}, 
\chi_{l2}^{(R_0)}, \chi_{l3}^{(R_0)}, \chi_{l4}^{(R_0)}$. The expressions of
$A_{l1}^{(R_0)}, A_{l2}^{(R_0)}$ are
\begin{equation}
A_{l1}^{(R_0)}=-\frac{J_{|l-\alpha|}^\prime (kR_0)-(\kappa/k)f_{l1}^{(R_0)}
J_{|l-\alpha|}(kR_0)}{H_{|l-\alpha|}^{(1)\prime} (kR_0)-
(\kappa/k)f_{l1}^{(R_0)}H_{|l-\alpha|}^{(1)}(kR_0)} ,
\end{equation}
\begin{equation}
A_{l2}^{(R_0)}=-\frac{J_{|l+1-\alpha|}^\prime (kR_0)-(\kappa/k)f_{l2}^{(R_0)}
J_{|l+1-\alpha|}(kR_0)}{H_{|l+1-\alpha|}^{(1)\prime} (kR_0)-
(\kappa/k)f_{l2}^{(R_0)}H_{|l+1-\alpha|}^{(1)}(kR_0)} ,
\end{equation}
where
\begin{equation}
f_{l1}^{(R_0)}=\frac{E+Mc^2}{E+Mc^2-U}
\Lambda_{l1}^{(R_0)}-\frac{U}{E+Mc^2-U}\frac{l-\alpha}{\kappa R_0},
\end{equation}
\begin{equation}
f_{l2}^{(R_0)}=\frac{E+Mc^2}{E+Mc^2-U}
\Lambda_{l2}^{(R_0)}+\frac{U}{E+Mc^2-U}\frac{l+1-\alpha}{\kappa R_0} .
\end{equation}
 The quantities $\Lambda_{l1}^{(R_0)},
\Lambda_{l2}^{(R_0)}$ appearing in eqs. (35), (36) are the logarithmic
derivatives of the radial eigenfunctions written in 
eqs. (27),(28) or (29),(30), evaluated at $r=R_0$,
\begin{equation}
\Lambda_{l1}^{(R_0)}=
\frac{d\chi_{l1}^{(R_0)}/\kappa dr}{\chi_{l1}^{(R_0)}},
\Lambda_{l2}^{(R_0)}=
\frac{d\chi_{l2}^{(R_0)}/\kappa dr}{\chi_{l2}^{(R_0)}} \:\, {\rm for} \:\,
 r\rightarrow R_0, r<R_0.
\end{equation}

Due to the exponential dependence of the eigenfunctions in the case 
$\kappa R_0\gg 1$, the quantities defined in eq. (37)
converge to 1, $\Lambda_{l1}^{(R_0)}\rightarrow 1,\Lambda_{l2}^{(R_0)}
\rightarrow 1. $
The denominators in eqs. (33), (34) then are
\begin{eqnarray}
\lefteqn{H_{|l-\alpha|}^{(1)\prime} (kR_0)-
(\kappa/k)f_{l1}^{(R_0)}H_{|l-\alpha|}^{(1)}(kR_0)} \\ \nonumber
 & &\sim\frac{1}{E+Mc^2-U}\left[(E+Mc^2)\left(-\frac{|l-\alpha|}{kR_0}-
\frac{\kappa}{k}\right)+U\frac{|l-\alpha|+l-\alpha}{kR_0}\right] ,
\end{eqnarray}
\begin{eqnarray}
\lefteqn{H_{|l+1-\alpha|}^{(1)\prime} (kR_0)-
(\kappa/k)f_{l2}^{(R_0)}H_{|l+1-\alpha|}^{(1)}(kR_0)}\nonumber\\
 & &\sim\frac{1}{E+Mc^2-U}\left[(E+Mc^2)\left(-\frac{|l+1-\alpha|}{kR_0}-
\frac{\kappa}{k}\right)+U\frac{|l+1-\alpha|-(l+1-\alpha)}{kR_0}\right] ,
\end{eqnarray}
and are different from 0 for all values of $l$.
From the fact that, for  $kR_0\ll 1$, $J_{|l-\alpha|}(kR_0) \sim (kR_0)
^{|l-\alpha|}, \\
H_{|l-\alpha|}^{(1)}(kR_0) \sim (kR_0)^{-|l-\alpha|}$,
it follows that
\begin{equation}
A_{l1}^{(R_0)} \sim (kR_0)^{2|l-\alpha|},\:\,A_{l2}^{(R_0)}
\sim (kR_0)^{2|l+1-\alpha|},
\end{equation}
so that the coefficients $A_{l1}^{(R_0)}, A_{l2}^{(R_0)}$ are 
vanishing in the limit $kR_0\rightarrow 0$. Then from eqs. 
(6),(7),(31),(32)
we obtain the Dirac eigenfunctions of an electron in the presence of a
{\it shielded} magnetic string as
\begin{equation}
\chi_{l1}^{(sh)}=a_{l1}J_{|l-\alpha|}(kr)
\end{equation}
\begin{equation}
\chi_{l2}^{(sh)}=a_{l2}J_{|l+1-\alpha|}(kr)
\end{equation}
\begin{equation}
\chi_{l3}^{(sh)}=-\frac{i\hbar c a_{l2}}{E+Mc^2}\left(\frac{d}{d
r}+\frac{l+1-\alpha}{r}\right)J_{|l+1-\alpha|}(kr) ,
\end{equation}
\begin{equation}
\chi_{l4}^{(sh)}=-\frac{i\hbar c a_{l1}}{E+Mc^2}\left(\frac{d}{d
r}-\frac{l-\alpha}{r}\right)J_{|l-\alpha|}(kr) ,
\end{equation}
for $r>R_0, R_0\rightarrow 0$. These eigenfunctions have been first written 
in eqs. (16) of
ref. \cite{5}, but the simplified mathematical deduction presented there
is not applicable for the eigenfunctions of angular momentum $l=[\alpha]$.\\

For a given $\alpha$, the ensemble of eigenfunctions for the shielded
magnetic string, $U\not= 0$, and the ensemble of eigenfunctions for the bare
magnetic string, $U$=0, differ only by the eigenfunction of angular momentum
$l=[\alpha]$, the eigenfunctions for all other values of $l$ being identical.
If $0<\alpha<1$, the radial components of the eigenfunction $l$=0 for the 
bare string are
\begin{equation}
\chi_{01}^{(b)}=b_{01}J_{-\alpha}(kr),
\end{equation}
\begin{equation}
\chi_{02}^{(b)}=b_{02}J_{1-\alpha}(kr),\nonumber\\ 
\end{equation}
\begin{equation}
\chi_{03}^{(b)}=-\frac{i\hbar ck b_{02}}{E+Mc^2}J_{-\alpha}(kr),
\end{equation}
\begin{equation}
\chi_{04}^{(b)}=\frac{i\hbar ck b_{01}}{E+Mc^2}J_{1-\alpha}(kr),
\end{equation}
For $0<\alpha<1$, the radial components of the eigenfunction $l$=0 for the 
shielded string are
\begin{equation}
\chi_{01}^{(sh)}=a_{01}J_\alpha (kr),  
\end{equation}
\begin{equation}
\chi_{02}^{(sh)}=a_{02}J_{1-\alpha}(kr),
\end{equation}
\begin{equation}
\chi_{03}^{(sh)}=-\frac{i\hbar ck a_{02}}{E+Mc^2}J_{-\alpha}(kr),
\end{equation}
\begin{equation}
\chi_{04}^{(sh)}=-\frac{i\hbar ck a_{01}}{E+Mc^2}J_{\alpha-1}(kr) .
\end{equation}
The definitions of the parameters $\alpha$ in \cite{4} and \cite{5} differ by
sign. The choice $\alpha=qF/2\pi\hbar$ is similar to standard electromagnetic
relations like ${\bf F}=q{\bf E}$.\\

Although it is
not possible that all four components of the Dirac wave function should vanish
for finite $kR_0$ \cite{2}, it can be seen from eqs. (41)-(44),(49)-(52)
that the
vanishing of the four components is possible in the limit $kR_0\ll 1$, except
for $l=[\alpha]$ when the $\chi_{[\alpha]3}$ and $\chi_{[\alpha]4}$
components are divergent at $r=0$.\\

The scattering of a plane wave by a shielded magnetic string will now be 
compared with the scattering of a plane wave by a bare magnetic string.
As shown in \cite{5}, the scattering of a plane wave by a {\it shielded}
magnetic string can be represented, for $0<\alpha<1$, by the Dirac wave 
function  
\begin{equation}
\Psi_\alpha^{(sh)}=\left( \begin{array}{c} a_1\\ a_2\\ -\frac{\hbar c k a_2}
{E+Mc^2}\\ -\frac{\hbar c k a_1}{E+Mc^2}\end{array} \right)\psi_\alpha^{(sh)}
(r,\theta)e^{-iEt/\hbar}+
\left( \begin{array}{c} 0\\ 0\\ 
-\frac{i\hbar cka_2}{E+Mc^2}
\exp(i\pi\alpha/2)\sin(\pi\alpha)H_\alpha^{(1)}(kr)\\ 
\frac{\hbar c k a_1}{E+Mc^2}\exp(-i\pi\alpha/2)\sin(\pi\alpha)
H_{1-\alpha}^{(1)}(kr)e^{i\theta}\\ 
\end{array} \right) e^{-iEt/\hbar} ,
\end{equation}
where
\begin{equation}
\psi_\alpha^{(sh)}(r,\theta)=\sum_{l=-\infty}^{\infty}
e^{-\frac{i\pi}{2}|l-\alpha|}J_{|l-\alpha|}(kr)e^{il\theta}
\end{equation}
is the Aharonov-Bohm wave function for the scattering of spinless electrons 
\cite{1}.
Unlike $\psi_\alpha^{(sh)}$, which is vanishing as $r\rightarrow 0$, the
second term on the right-hand side of eq. (53) is divergent at $r$=0.
This term vanishes however in the  limit $\hbar k/Mc\rightarrow 0$.
The asymptotic expansion of the wave function, eq. (53), is
\begin{equation}
\Psi_\alpha^{(sh)}=\left( \begin{array}{c} a_1\\ a_2\\ -\frac{\hbar c k a_2}
{E+Mc^2}\\ -\frac{\hbar c k a_1}{E+Mc^2}\end{array} \right)
e^{-ikr\cos\theta+i\alpha\theta-iEt/\hbar}
-\left( \begin{array}{c} a_1e^{i\theta/2}\\ a_2e^{i\theta/2}\\ 
\frac{\hbar c k a_2}
{E+Mc^2}e^{-i\theta/2}\\ \frac{\hbar c k a_1}{E+Mc^2}e^{3i\theta/2}
\end{array} \right)\frac{\sin(\pi\alpha)}{\cos(\theta/2)}
\frac{e^{ikr+i\pi/4-iEt/\hbar}}{(2\pi kr)^{1/2}} .
\end{equation}

The scattering of a plane wave by a {\it bare} magnetic string
can be represented, for $0<\alpha<1$, by the Dirac wave 
function
\begin{equation}
\Psi_\alpha^{(b)}=\Psi_\alpha^{(sh)}+
\left( \begin{array}{c} ia_1e^{i\pi\alpha/2}
\sin(\pi\alpha) H_\alpha^{(1)}\\ 0\\ 0\\ \frac{\hbar c ka_1}{E+Mc^2}e^{i\pi 
\alpha/2}\sin(\pi\alpha)H_{\alpha-1}^{(1)}e^{i\theta}\end{array}\right)
e^{-iEt/\hbar} .
\end{equation}
The expression of the first component of the wave function in eq. (56) is
\begin{equation}
\psi_\alpha^{(b)}(r,\theta)=\sum_{l=-\infty,l\not= 0}^\infty  
e^{-\frac{i\pi}{2}|l-\alpha|}J_{|l-\alpha|}(kr)e^{il\theta}
+e^{\frac{i\pi\alpha}{2}}J_{-\alpha} (kr) .
\end{equation}
The asymptotic expansion of the wave function, eq. (56), is
\begin{equation}
\Psi_\alpha^{(b)}=\left( \begin{array}{c} a_1\\ a_2\\ -\frac{\hbar c k a_2}
{E+Mc^2}\\ -\frac{\hbar c k a_1}{E+Mc^2}\end{array} \right)
e^{-ikr\cos\theta+i\alpha\theta-iEt/\hbar}
+\left( \begin{array}{c} a_1e^{-i\theta/2}\\ -a_2e^{i\theta/2}\\ 
-\frac{\hbar c k a_2}{E+Mc^2}e^{-i\theta/2}\\ 
\frac{\hbar c k a_1}{E+Mc^2}e^{i\theta/2}
\end{array} \right)\frac{\sin(\pi\alpha)}{\cos(\theta/2)}
\frac{e^{ikr+i\pi/4-iEt/\hbar}}{(2\pi kr)^{1/2}} .
\end{equation}
According to eq. (53), the scattering of a plane wave by a shielded magnetic
string can be represented in the limit $\hbar k/Mc\rightarrow 0$ with the aid
of the Aharonov-Bohm wave function for spinless electrons, eq. (54). On the
other hand, according to eq. (56), there is a difference between
the scattering of polarized electrons by bare and shielded  strings, and the
difference persists in the non-relativistic limit.\\

Recently, Moroz \cite{6} remarked that the divergent terms in the eigenfunction
of an electron endowed with spin in the presence of a bare magnetic string are
also obtained by using the Pauli equation for an electron gyromagnetic ratio
$g$=2, while the divergent terms disappear if the anomalous part $g_m-2$=0.0023
of the gyromagnetic ratio, due to radiative corrections, is included in the
analysis. Therefore, for magnetic fluxes which are odd multiples of $h/2e$, 
the modifications to an eigenfunction described by Hagen \cite{4} will be 
observed only when $g_m-2<kr_0\ll 1$, i.e. when the tube of flux of radius 
$r_0$ can be
regarded as a string, and at the same time the tube is not so thin that the
anomalous magnetic moment wipes out the effect. As mentioned previosly, the 
radius of a tube of flux of $h/2e$ obtained with a 2 T magnetic field is
$r_{h/2e}=1.81\cdot10^{-8}$ m, so that if we choose $kr_{h/2e}$=0.2, then
the required kinetic energy of the incident electron would be 
$4.65\cdot10^{-6}$ eV, which is rather low for an electron interference
experiment.\\

An alternative approach by which the wave function for an electron 
interference experiment can be made very small in the region of the tube of 
magnetic flux is to represent the state of an incident electron by a wave
{\it packet} \cite{7}. The propagation of this wave packet can be studied
with the aid of a Green's function \cite{3},\cite{8}. If relativistic 
corrections are neglected, the state of the incident electron is described
by the two principal components of the Dirac wave function, and
for a tube of magnetic flux oriented in the z direction these principal
components are solutions of decoupled equations of the Schr\"{o}dinger type.
Then for each of the two principal components, the Green's function can be
obtained as a sum of products of eigenfunctions. It was found that
the eigenfunctions for a bare magnetic string and for a shielded magnetic
string differ only for $l=[\alpha]$. Thus, if $\alpha>0$, the wave function
$J_{\alpha-[\alpha]}(kr)\exp(i[\alpha]\theta)$ would be replaced by
$J_{[\alpha]-\alpha}(kr)\exp(i[\alpha]\theta)$. The difference between the
Green's function for the bare magnetic string $G_\alpha^{(b)}$ and the 
conventional Green's function $G_\alpha$ \cite{3} is, in the non-relativistic
limit, 
\begin{equation}
G_\alpha^{(b)}-G_\alpha=\frac{1}{2\pi}\int_0^\infty k dk
\left(J_{[\alpha]-\alpha}(kr)
J_{[\alpha]-\alpha}(kr^\prime)-J_{\alpha-[\alpha]}(kr)
J_{\alpha-[\alpha]}(kr^\prime)\right)e^{-i\hbar k^2t/2M+i[\alpha]
(\theta-\theta^\prime)} .
\end{equation}
The result of the integration is \cite{9}
\begin{eqnarray}
\lefteqn{G_\alpha^{(b)}-G_\alpha=-\frac{iM}{2\pi\hbar t}
\left[e^{-\frac{i\pi}{2}([\alpha]-\alpha)}J_{[\alpha]-\alpha}\left(
\frac{Mrr^\prime}{\hbar t}\right)-e^{-\frac{i\pi}{2}(\alpha-[\alpha])}
J_{\alpha-[\alpha]}\left(\frac{Mrr^\prime}{\hbar t}\right)\right]}\nonumber\\
 & &\times\exp\left(\frac{iM}{2\hbar t}(r^2+r^{\prime 2})
+i[\alpha](\theta-\theta^\prime)\right) ,
\end{eqnarray}
which can also be written as
\begin{equation}
G_\alpha^{(b)}-G_\alpha=\frac{M}{2\pi\hbar t}\sin\left(\pi(\alpha-[\alpha])
\right)
e^{\frac{i\pi}{2}(\alpha-[\alpha])}
H_{\alpha-[\alpha]}^{(1)}\left(\frac{Mrr^\prime}{\hbar t}\right)
\exp\left(\frac{iM}{2\hbar t}(r^2+r^{\prime 2})
+i[\alpha](\theta-\theta^\prime)\right) .
\end{equation}
For $Mrr^\prime/\hbar t \gg 1$ the difference $G_\alpha^{(b)}-G_\alpha$ is
\begin{equation}
G_\alpha^{(b)}-G_\alpha=\left(\frac{M}{2\pi^3\hbar t rr^\prime}\right)^{1/2}
\sin\left(\pi(\alpha-[\alpha])\right)\exp\left[\frac{iM}{2\hbar t}
(r^2+r^{\prime 2}+2rr^\prime)+i[\alpha](\theta-\theta^\prime)-\frac{i\pi}{4}
\right]
\end{equation}

The initial wave function of a packet of width $\delta$, centered at the time
$t$=0 at the point $\rho_0, \theta_0$ and of incident momentum $\hbar k$ 
oriented
in the $-$x direction can be represented as \cite{3}
\begin{equation}
\Psi_{\delta,\alpha}(r^\prime,\theta^\prime,0)=\frac{1}{\pi^{1/2}\delta}
\exp\left[i\alpha\theta^\prime
-ikr^\prime\left(1-\frac{\theta^{\prime 2}}{2}\right)
-\frac{r^{\prime 2}+\rho_0^2-2r^\prime \rho_0
\left[1-(\theta^\prime-\theta_0)^2/2\right]}{2\delta^2}\right] .
\end{equation}
The plane-wave contribution
$-ikr^\prime \cos \theta^\prime $ could be expanded about $\theta^\prime=0$
because $|\theta_0|\ll 1, \delta/\rho_0\ll 1.$
The difference at the point of polar coordinates $r, \theta$ and at the time
$t$ of the wave 
functions representing the scattering by the bare string and by the shielded 
string can be obtained as
\begin{equation}
\Delta(r,\theta,t)=\int\left(G_\alpha^{(b)}-G_\alpha\right)
\Psi_{\delta,\alpha}r^\prime dr^\prime d\theta^\prime .
\end{equation}
For $k\rho_0\gg 1, kr\gg 1, \rho_0\ll k\delta^2, r\ll k\delta^2$, the 
difference
$\Delta$ is
\begin{equation}
\Delta(r,\theta,t)=\frac{e^{i\pi/4}}{2^{1/2}\pi\delta}
\sin\left(\pi(\alpha-[\alpha])\right)
\frac{\exp(ikr+i[\alpha]\theta)}{(kr)^{1/2}}\exp\left[-\frac{i\hbar k^2t}{2M}
-\frac{\rho_0^2\theta_0^2}{2\delta^2}-\frac{\left(r+\rho_0-\frac{\hbar kt}{M}-
\frac{\rho_0\theta_0^2}{2}\right)^2}{2\delta^2}\right] ,
\end{equation}
where $\alpha>0$. For times $t$ such that $t>\rho_0(1-\theta_0^2/2)M/\hbar k$,
the difference $\Delta$ of the wave functions describing the scattering of the
incident packet by a bare string and by a shielded string is proportional
to $\exp(-\rho_0^2\theta_0^2/2\delta^2),$ or $\exp(-d^2/2\delta^2)$, where
$d=\rho_0\theta_0$ is the impact parameter of the incident wave packet. As was
shown in \cite{7}, in the case of the scatering of a wave {\it packet}
by a magnetic string the whole scattering amplitude is proportional to the
exponentially small factor $\exp(-d^2/2\delta^2)$, and we have just seen that
the replacement of $J_{\alpha-[\alpha]}$ by $J_{[\alpha]-\alpha}$ does not 
change this fact.\\ 

The scattering by a shielded magnetic string and the scattering
by a bare magnetic string are really different only when there is a significant
probability density in the region of the magnetic flux, i.e. when there is
a direct interaction between the magnetic moment of the electron and the
magnetic {\it field}. 
This is basically due to the fact that the interaction with a magnetic
dipole moment is proportional to the magnetic {\it field}, and for a shielded
string the magnetic field is equal to zero in the region where there is
a non-vanishing probability density. Actually, as shown recently \cite{10},
the Aharonov-Bohm effect is relevant not so much for the problem of the
description of the electromagnetic continuum by field strengths or
electromagnetic potentials, but rather it demonstrates the global character of
the states in quantum mechanics. \\

ACKNOWLEDGMENT\\

This work has been supported by a research grant from the Romanian Academy
of Sciences.\\


\newpage
\begin{thebibliography}{10}
\bibitem{1} Y. Aharonov and D. Bohm, Phys. Rev.  115  (1959) 485.
\bibitem{2} U. Percoco and V. M. Villalba, Phys. Lett. A  140  (1989) 105.
\bibitem{3} S. Olariu and I. I. Popescu, Rev. Mod. Phys.  57  (1985) 339.
\bibitem{4} C. R. Hagen, Phys. Rev. Lett.  64  (1990) 503.
\bibitem{5} S. Olariu, Phys. Lett. A  144  (1990) 287.
\bibitem{5a} V. Berestetski, E. Lifchitz and L. Pitayevski, Th\'{e}orie
quantique relativiste, Premi\`{e}re partie, in L. Landau and E. Lifchitz,
Physique th\'{e}orique, Tome IV (Mir, Moscou, 1972), pp. 95, 100.
\bibitem{5b} Ph. M. Morse and H. Feshbach, Methods of theoretical physics
(McGraw-Hill, New York, 1953) p. 604.
\bibitem{5c} Ref. \cite{5b}, pp. 623-624.
\bibitem{6} A. Moroz, Phys. Rev. A  53  (1996) 669.
\bibitem{7} S. Olariu and I. I. Popescu, Phys. Rev. D  27 (1983)  383.
\bibitem{8} M. Kretzschmar, Z. Physik  185  (1965) 84.
\bibitem{9} G. N. Watson,  A treatise on the theory of Bessel functions
(Cambridge University Press, Second Edition, 1948), pp. 77 and 395.
\bibitem{10} S. Olariu, Phys. Rev. A, accepted for publication.
\end{thebibliography}
\end{document}